\documentstyle[epsfig,multicol,aps,prl]{revtex} 
\renewcommand{\narrowtext}{\begin{multicols}{2}  
\global\columnwidth20.5pc}  
\renewcommand{\widetext}{\end{multicols}  
\global\columnwidth42.5pc}    
\newcommand{\eq}{\begin{equation}}
\newcommand{\ee}{\end{equation}}
\newcommand{\eqa}{\begin{eqnarray}}
\newcommand{\eea}{\end{eqnarray}}

\begin{document}
\draft
\title{Role of disorder in half-filled high Landau levels}

\author{D.~N. Sheng$^a$, Ziqiang Wang$^b$, and B. Friedman$^c$} 
\address{$^a$Department of Physics and Astronomy, California State
University, Northridge, CA 91330}
\address{$^b$Department of Physics, Boston College, Chestnut Hill, MA 02467}
\address{$^c$Department of Physics, Sam Houston State University,
Huntsville, Texas 77341-2267}

\date{\today}
\maketitle
\begin{abstract}
We study the effects of disorder on the quantum Hall stripe phases
in half-filled high Landau levels using exact numerical diagonalization.
We show that, in the presence of weak disorder, a 
compressible, striped charge density wave, becomes the true ground state. 
The projected electron density profile resembles that of a smectic liquid.
With increasing disorder strength $W$, we find that there exists a
critical value, $W_c\sim0.12 e^2/\epsilon\ell$, where a transition/crossover 
to an isotropic phase with strong local electron density fluctuations 
takes place.  The many-body density of states
are qualitatively distinguishable in these two phases and help
elucidate the nature of the transition. 

\end{abstract}
\narrowtext

Recent experiments have discovered a set of new, compressible states 
with anisotropic low-temperature 
magnetotransport properties in high mobility quantum Hall systems 
near half-filling of the high Landau levels (LLs) \cite{experi}. 
Subsequently, it was proposed that these properties, emerging as
a consequence of strong long-range electronic correlations and the 
quenching of the kinetic energy, are manifestations of
the electronic liquid-crystal phases with partially broken translation
and rotational symmetries \cite{fradkin,mac}.
The anomalous anisotropic transport can be
interpreted as arising from a rotational symmetry breaking
ground states, i.e. a smectic or a stripe ordered charge-density wave (CDW)
phase \cite{fradkin,mac,fertig,hap,barci1}.
Remarkably, a unidirectional CDW state
has been predicted to occur in half-filled high LLs 
in the Hartree-Fock (HF) theory \cite{hfm}.
However, it has been known that the smectic liquid phase in the
HF theory is always unstable to translation-symmetry breaking
density modulations along the stripes \cite{fradkin,mac,fertig,hap,barci1}.
The latter produces an anisotropic but
incompressible striped crystal that is insulating in both directions,
inconsistent with experiments.  It is thus important to consider
quantum fluctuations beyond the HF theory, which is under
active investigation. 

In this paper, we study the effects of disorder
on the ground state and low energy excitations in half-filled high
LLs using exact numerical diagonalizations of
finite systems with up to twelve electrons.
Previously, exact diagonalizations were carried out 
by Rezayi, Haldane, and Yang \cite{rhy} (hereafter referred to as RHY)
for {\it clean} systems.
Although the ground state in this case does not directly exhibit 
stripe order, RHY discovered that 
the static density response function $\chi({\bf q})$ and the correlation
function $S_0 ({\bf q})$
are both sharply and strongly peaked at ${\bf q}^*=(q_x^*,0)$, 
indicative of the tendency towards the formation of a unidirectional CDW. 
They conjectured that in the thermodynamic limit, an external modulation with a
period matching ${\bf q}^*$ will result in a quantum Hall stripe phase.
We will show that the presence of disorder leads to several interesting and
important results for the ground state and the low energy excitations:
(i) A very weak {\it random}
potential associated with disorder induces a compressible
quantum smectic phase as the true ground state.
(ii) There exists a 
critical/characteristic disorder strength,
$W_c\sim0.12 e^2/\epsilon\ell$ where $\ell$ is the magnetic length,
which marks a transition/crossover from the anisotropic smectic phase at 
$W<W_c$ to an isotropic fluid phase at  $W> W_c$ \cite{hfmelting}.
We illustrate the changes in the collective excitation spectrum in
terms of the disorder-averaged {\it many-body} density of states (mDOS)
at low energies. We show that the latter exhibits an intriguing crossover 
from that resembles a finite density of in-gap states in the 
smectic phase to that of a vanishing Coulomb gap behavior in the isotropic
phase.

We consider a two-dimensional system in an $a\times b$ rectangular cell
with periodic boundary conditions imposed in both $x$ and $y$ directions
\cite{notebc}.  
In the presence of a strong magnetic field, one can project
the Hamiltonian onto the up-most, partially-filled, $N$-th Landau level.
The projected Hamiltonian in the presence of both Coulomb interaction 
and disorder can be written as:
\eqa
H&=&\sum _{i<j} \sum _{ \bf {q}}  e^{-q^2/2 } [L_N (q^2/2)]^2 V(q)
e^{i {\bf q} \cdot {\bf (R_i -R_j)} }
\nonumber \\
&+&\sum _{i} \sum _{ \bf {q}} e^{-q^2/4} L_N (q^2/2) V_{\rm imp}(q)
e^{i{\bf q }\cdot {\bf R_i }}
\label{h}
\eea
where ${\bf R}_i$ is the guiding center coordinate of the $i$-th 
electron \cite{rhy},
$L_N (x) $ is the Laguerre Polynomial,  $V(q)=2\pi e^2/\epsilon q$ is the
Coulomb potential, and $V_{\rm imp}(q)$ is the impurity potential.
We set $\ell=1$ and $e^2/\epsilon\ell=1$ for convenience.
In Eq.~(\ref{h}), the wave vector ${\bf q}=2\pi(i_x/a,i_y/b)$ 
with $i_x$ and $i_y$ integers.
The disorder potential is generated according to the correlation relation 
in $q$-space $<V_{\rm imp}(q)V_{\rm imp}(-q')>=W^2 \delta _{q,-q'}$,
which corresponds to $<V({\bf r})V({\bf r'})>=W^2  \delta (\bf {r-r'})$
in real space, where $W$ is the strength of the disorder in
units of $e^2/\epsilon\ell$.
We consider up to $N_e=12$ electrons, spanning a Hilbert space 
of size $N_{\rm basis}=2,704,156$.
Since the disorder potential spoils the symmetry classifications of
the states used in RHY, we must deal with the entire Hilbert space.
We obtain the exact low energy eigenstates and eigen wave-functions 
using the Lanczos diagonalization method.

\begin{figure}      
\vspace{-0.5truecm}  
\center      
\centerline{\epsfysize=2.7in      
\epsfbox{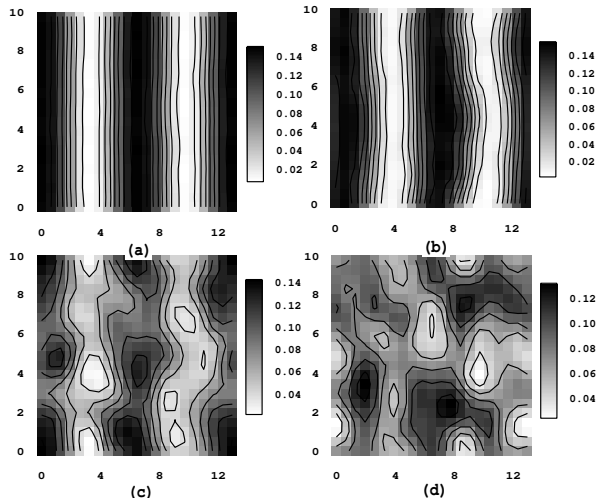}}  
\begin{minipage}[t]{8.1cm}     
\caption{
The projected electron density $\rho ({\bf r})$
at four different disorder strengths
for $N_e=10$, $N=2$ and aspect ratio $A=0.75$.
(a) $W=0.015$. (b) $W=0.045$. (c) $W=0.10$. (d) $W=0.14$.}
\label{fig1}
\end{minipage}
\end{figure}

The most direct probe of charge density order in the ground state
is the projected electron density $\rho({\bf r})$, i.e.
the equivalent electron density in the lowest LL 
describing the spatial distribution of the guiding center (GC) \cite{rhy,book}.
It is given by
\eq
\rho ({\bf r})=\sum _{j1,j2} \langle 0| a_{j_1}^+
 \phi ^*_{j_1} ({\bf r}) \phi _{j_2} ({\bf r})a_{j_2}|0 \rangle ,
\label{rho}
\ee
where the summation is over all single particle orbits and 
$\phi _j ({\bf r})$ is the corresponding wavefunction in the lowest LL.
In Fig.~1, we show how $\rho ({\bf r})$ in the ground state
evolves with disorder strength $W$ for
the number of electrons $N_e=10$ in the $N=2$ LL.
The aspect ratio of the system is $A=b/a=0.75$,
which is an optimized geometry for observing the instability toward the
formation of a stripe phase as suggested by RHY.
For a very weak disorder strength,
$W=0.015$, the typical behavior of
$\rho({\bf r})$ is shown in Fig.~1a
as a function of the two-dimensional coordinates.
Two nearly perfect stripes are formed along the $y$-direction
with essentially no density modulations along the stripes.
This is the first direct demonstration by exact method of the stripe 
ordered ground state in finite systems in the presence of weak disorder.

It is important to note that in the clean case studied by RHY,
the ground state and all exact many-body eigenstates are 
spatially {\it homogeneous} due to the translational symmetry
of the Hamiltonian. Only certain combinations of the five low-lying,
nearly degenerate eigenstates, 
separated by the unique wave vector ${\bf q^*}$, 
can give rise to a stripe phase that 
has a substantial overlap with the HF unidirectional CDW state. 
It may be possible to realize such a constructed stripe phase in the 
thermodynamic limit in clean systems if an external potential 
modulation is present with a wave vector that matches ${\bf q^*}$ \cite{rhy}.
However, when a relatively weak disorder is turned on, it causes a
natural mixing of the low-lying eigenstates and the near degeneracy
is lifted. At the same time, the random disorder potential necessarily
introduces an oscillatory component matching the wave vector ${\bf q}^*$.
The presence of the enormous
static density response $\chi({\bf q}\to
{\bf q}^*)$ then gives rise to the unidirectional CDW, or the stripe order
in the ground state that we observe.

\begin{figure}      
\vspace{-0.5truecm}  
\center      
\centerline{\epsfysize=1.7in      
\epsfbox{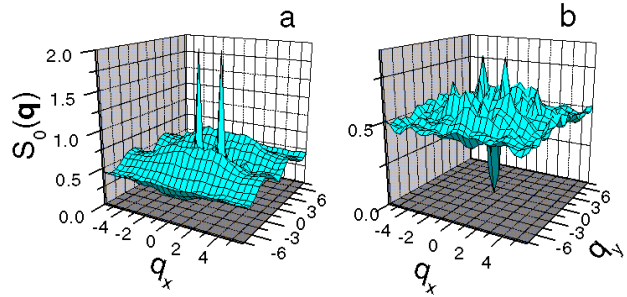}}  
\begin{minipage}[t]{8.1cm}     
\caption{
The static density-density correlation function
$S_0 ({\bf q})$ at (a) weak disorder $W=0.09 < W_c$ and
(b) strong disorder $W=0.14> W_c$.}
\label{fig2}
\end{minipage}
\end{figure}

To further support the picture of weak disorder induced stripe order
in the ground state, we calculated the projected equal time density-density 
correlation function $S_0 ({\bf q})$ defined as follows:
\eqa
S_0({\bf q})&=&\frac 1 {N_e} <0|\sum _{i,j}  e^{i{\bf q } 
\cdot {\bf (R_i-R_j) } }|0> .
\label{s0}
\eea
We find that for weak disorder of strength $W\le 0.05$, $S_0({\bf q})$
is quantitatively indistinguishable from the result of RHY for the 
corresponding clean system, showing strong and sharp peaks at
${\bf q}^*=(\pm 0.97,0)$ for $2\pi/a=0.485$, with a peak value 
$S_0 ({\bf q}^*)=3.7$ in the case of $N_e=10$.
The peak value of $S_0 ({\bf q}) $ increases with the number of
electrons to 4.9 at $N_e=12$. 
It can be seen from Fig.~1a that the width of the stripes  
is given by $D_s=a/2$, which equals $2\pi/q_x^*$ 
with $q_x=4\pi/a$, as implied in this picture. It is
interesting to note that the edges of the stripes are rather
soft, resembling that of a single dominant harmonic variation.
Next we discuss the evolution of the ground state with increasing
disorder strength $W$. Fig.~1b shows the real space profile
of  $\rho({\bf r})$ at $W=0.045$. There are
still two complete stripes with the same stripe width
$D_s$. Although density modulations along the stripes are
still minimal, the shape of the stripes exhibits pronounced
variations in the stripe direction. Thus we conclude that
for weak to moderate disorder strength, the ground state
is in the quantum Hall smectic phase\cite{fradkin,barci2,fogler}. 

The behavior of
the static density-density correlation function $S_0({\bf q})$
in the smectic phase is exemplified in Fig.~2a
for $W=0.09$. Despite the relatively strong disorder potential,
the peaks of $S_0 ({\bf q})$ are still locked at ${\bf q^*}$, but
the peak value has been reduced to $1.96$. 
The entire structure of $S_0({\bf q})$  remains qualitatively unchanged
from the clean limit \cite{rhy}. The absence of new, emerging peaks at other 
${\bf q}$ suggests that the smectic ground state 
is stable against density modulations along the stripes, even
in the presence of translational-symmetry breaking disorder potentials.

We next demonstrate that further increasing the disorder strength $W$ 
eventually makes the smectic phase unstable.
The projected density $\rho({\bf r})$ is shown in Fig.~1c for $W=0.1$.
At this value of $W$, the smectic phase is on the brink of being unstable.
$\rho({\bf r})$ shows very strong shape fluctuations and
significant density modulations along the stripes with 
detectable inter-stripe correlations.
At a still larger $W=0.14$, the $\rho({\bf r})$ plotted in Fig.~2d
shows that the long-range stripe order is lost: the stripes
become raptured and riddled with defects, very much like in the nematic phase.
Furthermore, the local orientation of the short-range stripes 
show indications that it has been rotated away from the $y$-axis.
The corresponding $S_0({\bf q})$ is
shown in Fig.~2b. It is clear that
the primary peaks around ${\bf q^*}$ become very much broadened with
significantly reduced peak values. Concomitantly, several competing peaks 
appear in $S_0 ({\bf q})$, marking the breakdown of the smectic
phase by random disorder and the emergence of a nematic/isotropic phase
with local electron density variations. The positions of these peaks
are random and dependent on the disorder configuration.
These results suggest that there exists a smectic to nematic-like
or isotropic phase transition, or a sharp crossover that takes place at a
critical/characteristic $W_c$ between
$W=0.10$ and $W=0.14$. We note that it is
somewhat difficult to distinguish between a nematic-like and an isotropic
phase with short-range stripe correlations in our finite size systems.

In order to pin down $W_c$, 40 different disorder configurations have been
studied in detail. We find that
the disorder strength $W_c$, at which multiple random peaks
develop in $S_0({\bf q})$ and compete with the
primary peaks at ${\bf q}^*$,
is always around a narrow region of $W_{c}=0.12$. As expected, the
same value of $W_c$ can be extracted directly from the spatial
distributions of $\rho({\bf r})$ shown in Fig.~1.
In Fig.~3a, we plot the disorder averaged peak values of
$S_0({\bf q}^*)$ 
for $N_e=8$ and $N_e=10$
as a function of $W$.
$S_0({\bf q^*})$ stays almost constant for relatively weak disorder
($W\le 0.05$). Then it begins to drop rather quickly around 
$W=0.07$, while the strong peaks of $S_0 ({\bf q} )$ 
remain at ${\bf q}^*$ and the emergent peaks at other values of ${\bf q}$
are absent. For even stronger disorder, $S_0 ({\bf q}^*)$ becomes
comparable to
the background value of $S_0 ({\bf q})$, which is approximately 0.5 for 
electrons at half-filling, and simultaneously
random peaks emerge around ${\bf q}^*$.
Such a change occurs around $W_c=0.12$ for $N_e=10$ as marked in Fig. 3a.
Due to the finite size effect, the value of $W_c$ for $N_e=8$ 
is around $0.1$. 
\begin{figure}[t1!]
\vspace{-3.0cm}
\epsfxsize=5.5 cm
\centerline{\epsffile{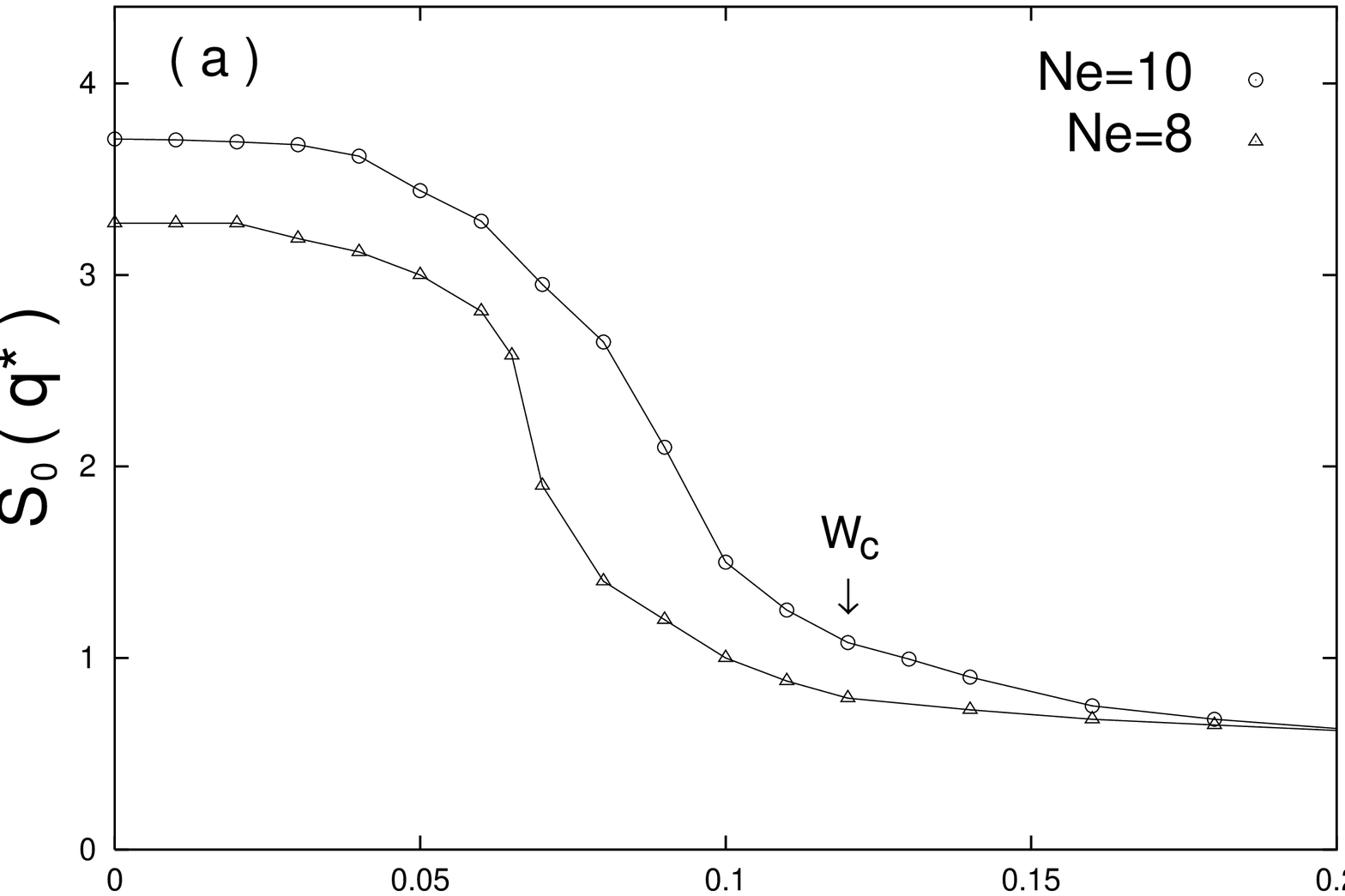}}
\vspace{-2.0cm}
\epsfxsize=5.5 cm
\centerline{\epsffile{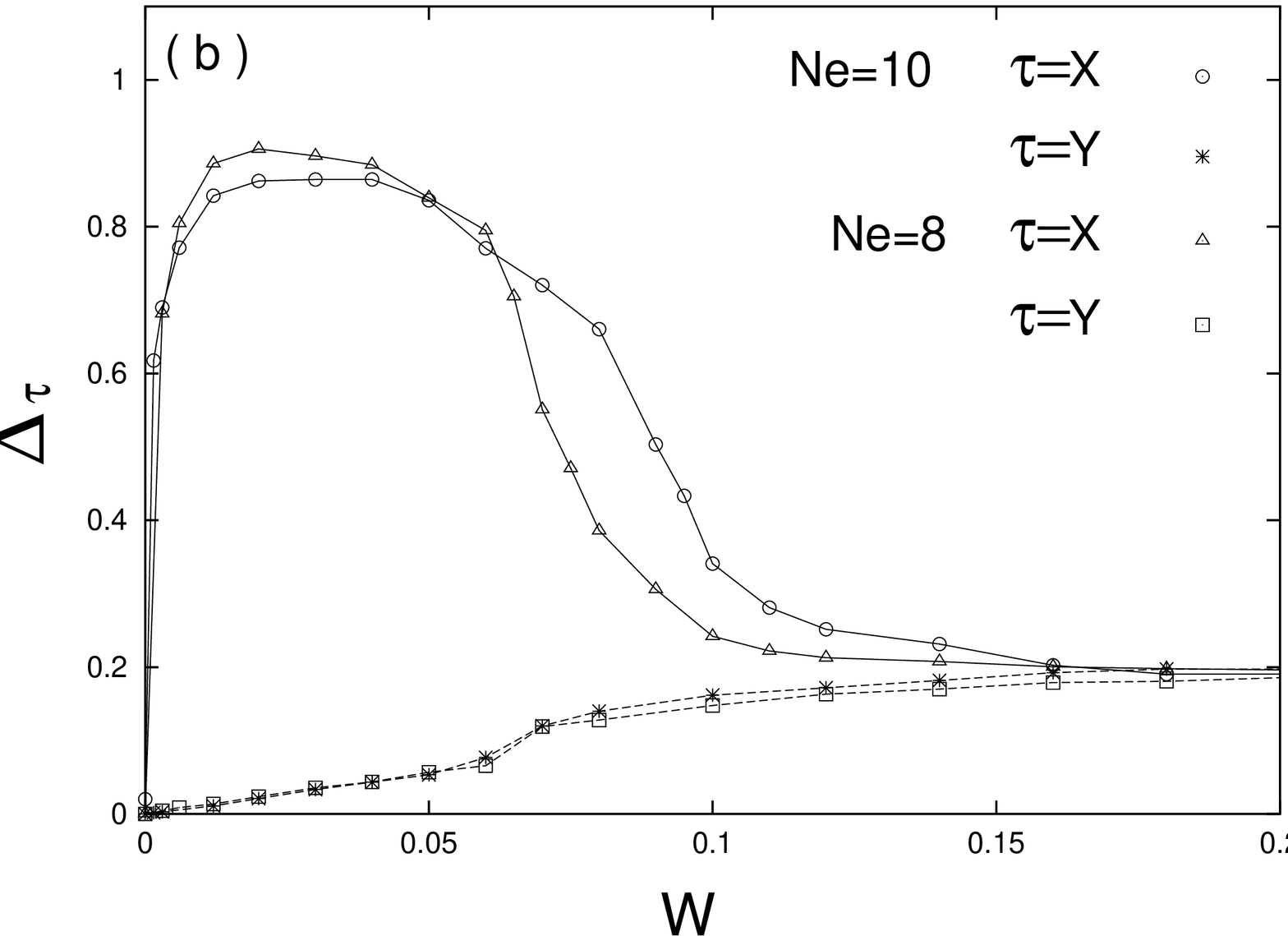}}
\vspace{0.9cm}
\begin{minipage}[t]{8.1cm}     
\caption{
(a). The peak value of the $S_0 ({\bf q})$
at the primary wave vectors ${\bf q}={\bf q}^*$ as
a function of  $W$ with a critical value
$W_c\simeq0.12$. (b) The constructed ``order parameters''
of the smectic phase $\Delta_x$ and $\Delta_y$ as a function
of  $W$.}
\label{fig3}
\end{minipage}
\end{figure}

We next construct the ``order parameters'' of the smectic phase
$\Delta _x $ and $\Delta _y$, which essentially describe 
the overlap of the true ground state with the HF unidirectional CDW
state \cite{rhy},
\eq
\Delta_x=<[\sum _{j_x} \frac 1 {n_s} (n_{j_x}- n_{{j_x}+j0})^2]^{1/2}>.
\label{delta}
\ee
Here, $<...>$ denotes the average over 200 and 40 disorder 
configurations for $N_e=8$ and  $N_e=10$ respectively.
In Eq.~(\ref{delta}), the summation is over all single-particle cyclotron 
orbits along $x$-direction with a density $n_{j_x}=<0|a^+_{j_x}a_{j_x}|0>$,
$n_s=2N_e$ is the number of orbits in each LL whereas
$j_0=(D_s/2)/(a/n_s)=n_s/4$ is the half stripe width in units of
orbit number. $\Delta_y $ can be similarly defined.

In Fig.~3b, both $\Delta_x$ and $\Delta_y$ are shown as a function of
the disorder strength $W$.  
At $W=0$, the ground state is in the uniform phase for
the translational invariance of the system, 
$\Delta_x=\Delta_y=0$.
However, when we switch on a weak disorder potential that stabilize
the stripes stacked in the $x$-direction, $\Delta_x $ becomes much
larger than $\Delta_y$ and is close to 1. Only for the HF wavefunction,
where the cyclotron orbits are either fully occupied or completely empty,
$\Delta_x^{\rm HF}=1$.
The deviation of $\Delta_x$ from 1 at small $W$ is due
to the smooth variation of $<0|a_j^+a_j|0>$ between high and
low occupation orbits, i.e. the softness of the edges shown in $\rho({\bf r})$.
This remains true for $N_e=12$.
Although larger system sizes
are needed to rule out the finite size effect, we point out that
the absence of the sharp edges is an important observation since
many of the recent theoretical descriptions of the smectic 
phase\cite{fradkin,mac,fertig,hap} share a common starting 
point of coupled one-dimensional
edge states along the stripes, which requires the presence of
sharp edges.

When $W$ is increased to $0.1$, $\Delta_x $ drops quickly to about 
$0.3$, corresponding to the development of the local 
density modulations along the stripes shown in Fig.~1c.
Again we see that $W_c\simeq0.12$
can be considered as a transition point between a smectic and an 
isotropic phase
with comparable $\Delta_x$ and $\Delta_y$, signaling 
the loss of long-range anisotropic charge ordering in the ground state.
In the strong $W$ region, both $\Delta_x$ and $\Delta_y$ effectively describe
the strong local charge density fluctuations.
%
\begin{figure}[t3!]
\epsfxsize=5.5 cm
\vspace{-30mm}
\centerline{\epsffile{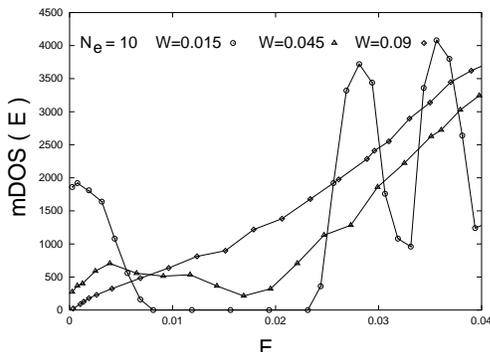}}
\vspace{10mm}
\begin{minipage}[t]{8.1cm}     
\caption{
The mDOS verses energy E at three different values of $W$
for  $N_e=10$ and $N=2$.}
\label{fig4}
\end{minipage}
\end{figure}

To develop further insights into the physical origin and the
nature of such a smectic, quantum Hall liquid  state, we 
plot, in Fig.~4, the many-body density of states (mDOS) for $N_e=10$
in the second LL.
It is clear that for weak disorder, represented by $W=0.015$,
the mDOS exhibits features of a finite density of in-gap states.
Specifically, there is a finite density of low-lying states separated from 
all excited states by a finite energy gap. This is a direct evidence that
the ground state is compressible. The gap value is $E_g\simeq0.015$ for 
both $N_e=10$ and $N_e=8$. 
All states below the gap (its number equals $N_e$)
have a similar $S_0({\bf q})$ and projected 
electron density $\rho ({\bf r})$,
each represents a stripe state in the $y$-direction 
shifted away by one orbit distance. They correspond
to long-wavelength phonon excitations. 
The higher energy states above the gap are isotropic,
corresponding to interstitial and defect-like excitations
accommodated by the impurity potential. The presence of such a finite
energy gap in weak disorder and the nature of the in-gap states suggest that
the smectic state obtained here is likely to be the
stable ground state in the thermodynamic limit.

Upon increasing $W$, the isotropic 
states move toward lower energy and the gap is reduced. 
At $W=0.045$, the gap region has turned into a minimum in the mDOS.
However, the mDOS remains finite in the limit $E\rightarrow 0$, although 
its value has been reduced as states are transfered to higher energies
by disorder.  The filling of the energy gap begins to have an impact
on the projected electron density in the ground state shown in Fig.~1b in 
the form of pronounced shape fluctuations along the stripe directions.
With the further increase of $W$, the gap eventually collapses.
At $W=0.09$, the mDOS as a function of $E$ exhibits a remarkable
behavior: it vanishes linearly in the limit 
$E\rightarrow 0$. This behavior remains qualitatively unchanged
in the strong $W$ region and is insensitive
to sample sizes for $N_e=8,10$ and $12$.
Further increase of $W$ in the absence of the spectral gap causes
strong mixing of the low-lying and the excited states and eventually 
destroy the smectic phase in favor of an isotropic phase with
short-range stripe correlations.

We have also studied the transport property of the anisotropic CDW state
by calculating the Thouless energy $\Delta E _{xx}$ or $\Delta E_{yy}$, 
which is the change in the ground state energy under the change of the 
boundary condition in the $x$ or $y$ direction from periodic to antiperiodic. 
We find that for weak disorder, e.g. $W=0.015$, $\Delta E_{yy}$
is about 50 times bigger than $\Delta E_{xx} $, suggesting \cite{note} a 
large transport anisotropy associated with the quantum Hall smectic phase
consistent with the experimental findings.

We would like to acknowledge helpful
discussions M. Fogler, E. Rezayi and M. P. A. Fisher. D.N.S. would like to 
thank the hospitality of the Aspen Center for Physics where this work was 
initiated.  D.N.S. is supported by Petroleum Research Fund grant No. 
ACS-PRF \# 36965-AC5 and Research Corporation award No. CC5643.
Z.W. is supported by DOE grant No. DE-FG02-99ER45747
and an award from Research Corporation.

\end{multicols} 

\begin{references}
\bibitem{note}  
For a noninteracting system, Thouless energy $\Delta E _{xx}$ or $\Delta 
E_{yy}$ can be related to the conductance quantitatively. For an interacting 
system, $\sigma _{xx}$ should still be a monotonic function of $\Delta 
E_{xx}$ as it measures the general extensiveness of the wavefunction by its 
sensitivity to the change in the boundary condition.
\end{references}
\end{document}